\documentclass[conference]{IEEEtran}
\IEEEoverridecommandlockouts
\usepackage{cite}
\usepackage{amsmath,amssymb,amsfonts}
\usepackage{algorithmic}
\usepackage{graphicx}
\usepackage{textcomp}
\usepackage{multirow}
\usepackage{tabularx}
\usepackage[table,xcdraw]{xcolor}
\usepackage[normalem]{ulem}
\useunder{\uline}{\ul}{}
\def\BibTeX{{\rm B\kern-.05em{\sc i\kern-.025em b}\kern-.08em
    T\kern-.1667em\lower.7ex\hbox{E}\kern-.125emX}}
\makeatletter
\newcommand{\linebreakand}{%
  \end{@IEEEauthorhalign}
  \hfill\mbox{}\par
  \mbox{}\hfill\begin{@IEEEauthorhalign}
}
\makeatother
    
\begin{document}

\title{Energy Consumption in RES-aware 5G Networks
}

\author{\IEEEauthorblockN{Adam Samorzewski}
\IEEEauthorblockA{
\textit{Poznan University of Technology} \\
Poznan, Poland \\
adam.samorzewski@doctorate.put.poznan.pl}
\and
\IEEEauthorblockN{Margot Deruyck}
\IEEEauthorblockA{
\textit{Ghent University -- IMEC}\\
Ghent, Belgium \\
margot.deruyck@ugent.be}
\and
\IEEEauthorblockN{Adrian Kliks}
\IEEEauthorblockA{
\textit{Poznan University of Technology} \\
Poznan, Poland \\
adrian.kliks@put.poznan.pl}
}

\maketitle

\begin{abstract}
In this work, the impact of using Renewable Energy Source (RES) generators in next-generation (5G) cellular systems on total power consumption (PC) has been investigated. The paper highlights the gain related to the use of photovoltaic (PV) panels and wind turbines (WTs) in the form of two factors -- the average extension of battery lifetime (AEBL) powering a single network cell and the average reduction in energy consumption (AREC) within the whole network. The examination has been conducted for four different seasons of the year and various configurations of available power sources. The provided system scenario was based on real data on weather conditions, building placement, and implemented mobile networks for the city of Poznan in Poland. Used RES generators were designed in accordance with the specifications of real devices.\footnote{Copyright © 2023 IEEE. Personal use is permitted. For any other purposes, permission must be obtained from the IEEE by emailing pubs-permissions@ieee.org. This is the author’s version of an article that has been published in the proceedings of the 2023 IEEE Global Communications Conference (GLOBECOM) by the IEEE. Changes were made to this version by the publisher before publication, the final version of the record is available at: https://dx.doi.org/10.1109/GLOBECOM54140.2023.10437451. To cite the paper use: A. Samorzewski, M. Deruyck and A.~Kliks, “Energy Consumption in RES-Aware 5G Networks,” \textit{GLOBECOM 2023 -- 2023 IEEE Global Communications Conference}, Kuala Lumpur, Malaysia, 2023, pp.~1024--1029, doi: 10.1109/GLOBECOM54140.2023.10437451 or visit https://ieeexplore.ieee.org/document/10437451.}
\end{abstract}


\section{Introduction}
\label{secIntroduction}

Today, the ICT sector is still responsible for huge emissions, which are systematically underestimated and could actually be as high as $2.1$ to $3.9$\% of the global GHG (Green House Gas) emissions~\cite{Freitag}. Networks -- both wired and wireless -- contribute about $25$\% to this value. Within the wireless network, the base station (BS) is the largest consumer. Even though the hardware is becoming more energy-efficient, a $5$G BS consumes about $4$ times more energy compared to a $4$G one~\cite{I}. A $5$G BS does provide a $16$ times higher throughput, but the 5G network will be more dense, increasing the network's energy consumption by $150$ to $170$\% by $2026$~\cite{I}. One way to deal with the impact of this rise in energy consumption is the use of renewable energy sources (RESs) such as e.g., solar, wind, water, or geothermal energy. By using RESs, we can not only reduce emissions but also protect our fossil fuels from future depletion. Unfortunately, due to the fluctuating provisioning of some RESs (because of varying weather conditions), their use in wireless networks has not been widely investigated yet. 
In this paper, we build upon the work of~\cite{Deruyck} and \cite{Castellanos} by investigating the use of solar and wind energy for a large-scale $5$G network in the city of Poznan, Poland. 
The novelty of the study lies in: suitable power consumption models for $5$G and its features like massive MIMO (Multiple Input Multiple Output) are now included for various frequencies $\left(800, 2100, 3500 \text{ MHz}\right)$; weather prediction models based on real historical data for the city of Poznan, Poland; and (dis)charging and transportation losses are considered for four scenarios accounting for solar or wind energy or both.

\section{Methodology}
\label{secMethodology}
\subsection{Network Design}
\label{subsecNetworkDesign}
For this study, the network planner discussed in~\cite{Castellanos} is further extended. This planner called GRAND (Green Radio Access Network Design) is a deployment tool that designs and optimizes wireless outdoor access networks toward power consumption and/or human exposure by selecting the optimal base station locations and settings based on the instantaneous bit rate request of the users active in the environment. The tool takes various inputs such as a shape file describing the $3$D environment for which the network will be optimized, a list of possible base station locations, and a list of active users, their locations, and their bit rate requests. In the first step, the tool makes a list of potential base stations to connect each user with (based on the user's bit rate request and the maximum allowable path loss). Once this graph of base station-user associations is made, a MIP (Mixed Integer Programming) solver is applied to find the optimal solution. To this end, an objective function is defined based on the envisioned optimization of power consumption, human exposure, or both. 

\subsection{Renewable Energy Sources}
\label{subsecRES}
In order to reduce the use of energy from the conventional electrical grid, it was considered to power the wireless network with PV panels and wind turbines. Furthermore, to reduce the loss of harvested resources, each base station has been equipped with a battery system to store the excess energy. The processes of energy generation were modeled within the GRAND tool in accordance with the works contained in \cite{Björnson, Peng, HomerCalc, Carrillo, Voltacon, Omni, vCalc} and widely described in Section \ref{secEnergyModels}. In addition, the parameters for both types of RES generators as well as for the battery were implemented with respect to the specifications of real devices, which can be found in \cite{PvPanel, WindTurbine, Battery}. 

\section{Scenario}
\label{secScenario}
\begin{figure}[!t]
\centering
\includegraphics[width=0.475\textwidth]{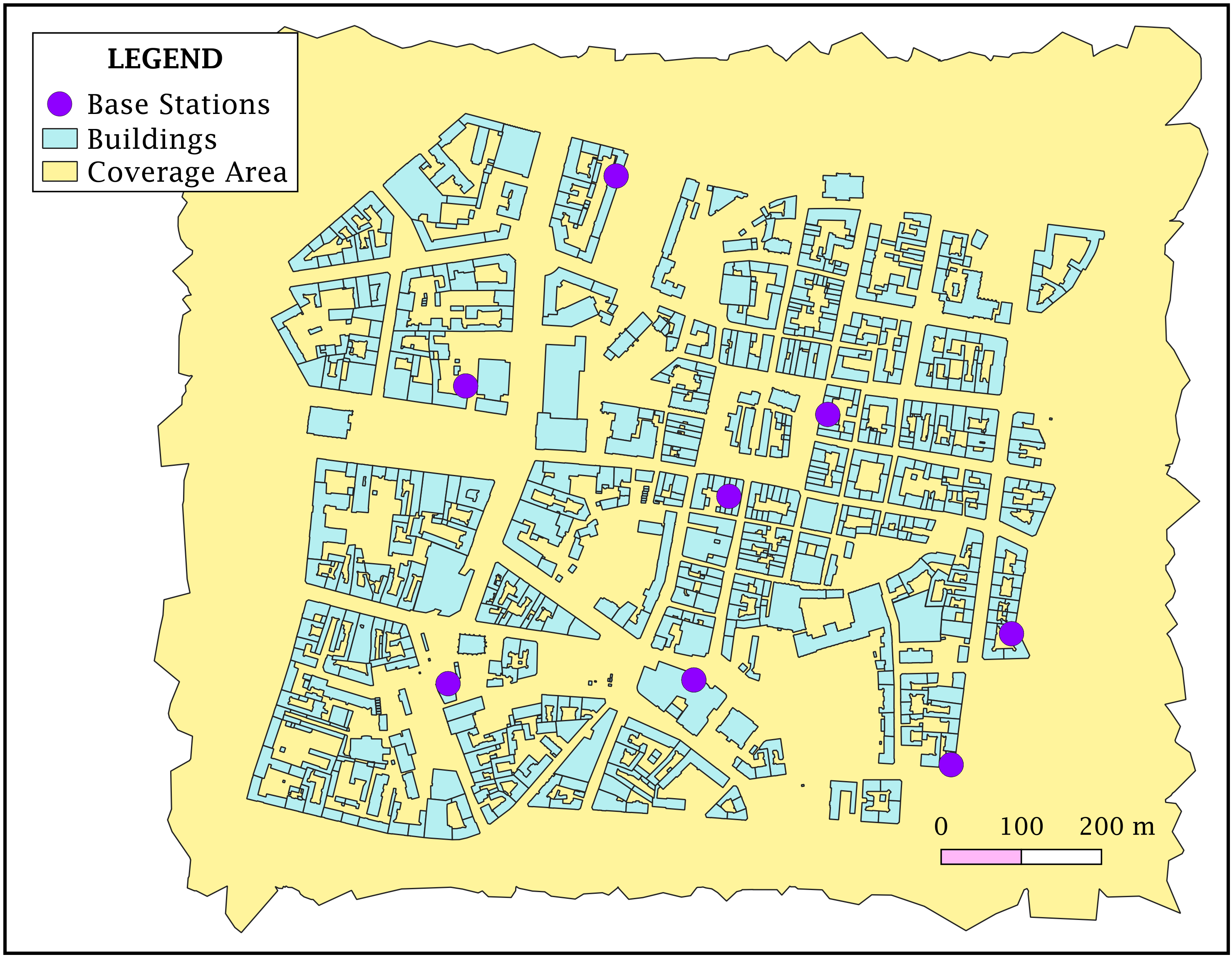}
\caption{Map of the examined area within the city of Poznan.}
\label{figScenario}
\end{figure}

In the examined scenario (illustrated in Fig.~\ref{figScenario}), the mobile network within the old market area of Poznan (Poland) was considered. In the very beginning, we assumed $8$ locations within the city, at which $5$G base stations are placed (designed on the basis of plans of the city of Poznan \cite{AreaData} and data of one Polish mobile operator given in \cite{NetworkData}). Each BS is mounted on one of the buildings and has $3$ cells using different frequency bands -- $800$, $2100$, and $3500$ MHz. In addition, cells configured in the third band were deployed to use Massive MIMO technology, which affects power consumption and signal propagation gain. For each hardware set of each cell type, there is a separate container (outside room) with air conditioners next to the BS tower. All the data about the placement of buildings and access nodes in the area as well as about the weather conditions were taken from the databases of historical records contained in \cite{NetworkData, AreaData, WeatherData}. Between buildings, there are $300$ outdoor users with dedicated equipment (UE), whose standings and throughput demands are fixed. All the locations of users are selected randomly within the simulator for each single run. Firstly, all the network cells are set to broadcast the radio signal with the maximum power. After distributing users within the examined area, the GRAND tool starts assigning them among the cells while optimizing cells' transmit powers and ensuring users' bit rate demands, simultaneously. Then, users begin exchanging data with the network continuously generating fixed traffic throughout the whole time of a simulation run. Within a single run, the calculations have been performed for four different dates (each from a different season of the year) to investigate the impact of time of the day and year on the energy harvesting characteristics of both RES types.

\section{Energy Models}
\label{secEnergyModels}
In this section, the mathematical formulas for modeling processes of energy consumption and production (prosumption) by the designed system have been presented. 

\subsection{MIMO Base Station}
\label{subsecMimoBs}
The model used to estimate the power consumption by $5$G base stations has been formulated in accordance with the work contained in \cite{Björnson}. In this contribution, from the perspective of energy cycle modeling, all the cells have been considered to be Massive MIMO (whether they are or not) using a different number of active antenna elements (AAEs), i.e., $5$G $800$ and $5$G $2100$ -- $1$ AAE, $5$G $3500$ -- $64$ AAEs. The mathematical formula that evaluates the total power consumption $\left(P_\text{MIMO}\right)$ by a single cell in the current time step $t$ is as below:
\begin{align}
    \label{eqPowerMIMO}
    P_\text{MIMO}\left(t\right)=P_\text{CP}\left(t\right)+P_\text{PA}\left(t\right)+\frac{P_\text{cool}\left(t\right)}{1-\sigma_\text{cool}}+P_\text{AUX}\left(t\right),
\end{align}
where $P_\text{PA}\left(t\right)=\frac{P_\text{TX}\left(t\right)}{\mu_\text{PA}}$ is the power consumed by the power amplifier. The parameters of $P_\text{TX}$ and $\mu_\text{PA}$ are the transmit power and efficiency of the amplifier, respectively. In turn, $\sigma_\text{cool}$ is the cooling loss factor. Next, the $P_\text{CP}$, $P_\text{cool}$, and $P_\text{AUX}$ are the powers spent by the transceiver, cooling, and auxiliary hardware components. The latter refers to additional equipment not related to providing the radio connection (e.g., lighting). The former was expressed as follows:
\begin{align}
    \label{eqPowerCP}
    \nonumber
    &P_\text{CP}\left(t\right)=P_\text{FIX}+P_\text{TC}\left(t\right)+P_\text{CE}\left(t\right)+P_\text{C/D}\left(t\right)+P_\text{BH}\left(t\right)\\&+P_\text{SP}\left(t\right),
\end{align}
where $P_\text{FIX}$ is the fixed power consumed by a cell node. The parameter of $P_\text{TC}\left(t\right)=M_\text{BS}\left(t\right) P_\text{CC}$ is the power utilized by the transceiver chains in the time step $t$, where $P_\text{CC}$ is the power that is required to run the circuit components (e.g. filters, I/Q mixers, etc.), and $M_\text{BS}$ is the number of presently active antenna elements of the cell. Next, $P_\text{CE}\left(t\right)=\frac{3B_\text{w}}{\tau_\text{c}\cdot \eta_\text{BS}}K_\text{UE}\left(t\right)\left(M_\text{BS}\left(t\right)\tau_\text{p}\left(t\right)+M_\text{BS}^{2}\left(t\right)\right)$ is the power needed by the channel estimators, which work according to the minimum mean-squared error (MMSE) scheme \cite{Björnson}. The parameters of $B_\text{w}$ and $\eta_\text{BS}$ are the channel bandwidth and computational efficiency, respectively. In addition, $\tau_\text{p}\left(t\right)=\text{RF}\cdot K_\text{UE}\left(t\right)$ is the number of samples allocated for pilots per coherence block in a specific time step $t$, where $\text{RF}$ is the pilot reuse factor and $K_\text{UE}$ is the current number of served users. In turn, $\tau_\text{c}=B_\text{c}\cdot t_\text{c}$ is the number of samples per coherence block, where $B_\text{c}$ and $t_\text{c}$ are the coherence bandwidth and time, respectively. Furthermore, 
$P_\text{C/D}\left(t\right)=P_\text{COD}\cdot\text{TR}_\text{DL}\left(t\right)+P_\text{DEC}\cdot\text{TR}_\text{UL}\left(t\right)$
is the total power consumed by a single cell of the BS for encoding $\left(P_\text{COD}\right)$ and decoding $\left(P_\text{DEC}\right)$ the information transferred through uplink, in short UL, $\left(\text{TR}_\text{UL}\right)$ and downlink, in short DL, $\left(\text{TR}_\text{DL}\right)$ connections in a particular time step $t$. The power model takes into account also the load-aware part of the consumption referring to the backhaul links -- $P_\text{BH}\left(t\right)=P_\text{BT}\big(\text{TR}_\text{UL}\left(t\right)+\text{TR}_\text{DL}\left(t\right)\big)$, where $P_\text{BT}$ is the backhaul traffic power. Finally, the power required by the network cell for operations related to signal processing (e.g., UL reception and DL transmission, computation of the combining/precoding vectors) compliant with the MMSE scheme $\left(P_\text{SP}\right)$ is as below: 
\begin{align}
    \label{eqPowerSP}
    &P_\text{SP}\left(t\right)=\frac{3B_\text{w}}{\tau_\text{c}\cdot \eta_\text{BS}}\Bigg[M_\text{BS}\left(t\right) K_\text{UE}\left(t\right) \big(\tau_\text{u}\left(t\right)+\tau_\text{d}\left(t\right)\big)\\ \nonumber &+\frac{\left(3M_\text{BS}\left(t\right)^{2}+M_\text{BS}\left(t\right)\right)K_\text{UE}\left(t\right)}{2}+\frac{M_\text{BS}\left(t\right)^{3}}{3}+2M_\text{BS}\left(t\right)\\ \nonumber &+M_\text{BS}\left(t\right)\tau_\text{p}\left(t\right)\big(\tau_\text{p}\left(t\right)-K_\text{UE}\left(t\right)\big)+M_\text{BS}\left(t\right) K_\text{UE}\left(t\right)\Bigg],
\end{align}
where $\tau_\text{d}\left(t\right)$ is the number of DL data samples per coherence block in the time step $t$. 

\begin{figure*}[!t]
    \begin{equation}
        \label{eqPowerCool}
        \resizebox{0.8\textwidth}{!}{$
        P_\text{cool}\left(t\right) =
        \begin{cases}
            P_\text{CP}\left(t\right)\cdot\sigma_\text{CP} - A_\text{s}\cdot\sigma_{s}\cdot\big|T_\text{s}-T_\text{a}\left(t, h_\text{BU}\right)\big|, & \text{if } T_\text{s} \geqslant T_\text{a}\left(t,h_\text{BU}\right) \text{ and } T_\text{s} - \frac{P_\text{CP}\left(t\right)\cdot\sigma_\text{CP}}{A_\text{s}\cdot\sigma_{s}} < T_\text{a}\left(t, h_\text{BU}\right) \\
            P_\text{CP}\left(t\right)\cdot\sigma_\text{CP} + A_\text{s}\cdot\sigma_{s}\cdot\big|T_\text{s}-T_\text{a}\left(t,h_\text{BU}\right)\big|, & \text{if } T_\text{s} < T_\text{a}\left(t,h_\text{BU}\right) \\
            0, & \text{otherwise}
        \end{cases}$}
    \end{equation}
\end{figure*}

The component $P_\text{cool}$ in Eq.~(\ref{eqPowerMIMO}) represents the power that is necessary to keep the adequate temperature $\left(T_\text{s}\right)$ inside the server rooms of the BS. The mathematical formula describing the power utilization for hardware cooling has been attached in Eq.~(\ref{eqPowerCool}), where $\sigma_\text{CP}$ and $T_\text{a}\left(t,h_\text{BU}\right)$ are the circuit heat transfer coefficient of the hardware placed inside the server room and the temperature of its outside ambient in the time step $t$, respectively. The parameter of $h_\text{BU}$ denotes the ground-relative altitude of the building, on which a particular BS (as well as its server rooms and PV panels) is placed. Next, $A_\text{s}$ and $\sigma_{s}$ are the surface area and heat transfer coefficient of a specific server room storing the equipment of the cell. The former is equal to $A_\text{s}=2 a_\text{s} b_\text{s}+2 a_\text{s} h_\text{s}+2 b_\text{s} h_\text{s}$, where $a_\text{s}$, $b_\text{s}$, and $h_\text{s}$ are the dimensions of the server room (container).

\subsection{PV Panel}
\label{subsecPvPanel}
There is a need to model the energy harvesting process performed by PV panels supplying each cell of each BS tower. The output power of the set of PV arrays $\left(P_\text{PV}\right)$ in a certain time step $t$ can be denoted as \cite{HomerCalc}: 
\begin{align}
    \label{eqPowerPV}
    P_\text{PV}\left(t\right)=N_\text{PV} P_\text{R,PV} f_\text{PV}\cdot\frac{\overline{G}_\text{T}\left(t\right)}{\overline{G}_\text{T,STC}}\bigg[1+\alpha_\text{P}\Big(T_\text{c}\left(t\right)-T_\text{c,STC}\Big)\bigg],
\end{align}
where $N_\text{PV}$, $P_\text{R,PV}$, and $f_\text{PV}$ are the total number of PV panels allocated per network cell, and the rated power and derating factor of a single one. In addition, the first one is the multiplication of the numbers of PV panels connected in series $\left(N_\text{PV,s}\right)$ and parallel $\left(N_\text{PV,p}\right)$. 
Next, $\overline{G}_\text{T}$ and $T_\text{c}$ are the parameters that denote the solar radiation incident on the PV array and its temperature. Thus, $\overline{G}_\text{T,STC}$ and $T_\text{c,STC}$ define the values of the same parameters but for standard test conditions (STC). Finally, $\alpha_\text{P}$ is the temperature coefficient of power dependent on the type of used PV panels. Besides, to assess the temperature of the PV cell $\left(T_\text{c}\right)$, the formula was used \cite{HomerCalc}:
\begin{align}
    \label{eqTc}
    &T_\text{c}\left(t\right)=\frac{T_\text{a}\left(t, h_\text{PV}\right)}{1+\left(T_\text{c,NOCT}-T_\text{a,NOCT}\right)\left(\frac{\overline{G}_\text{T}\left(t\right)}{\overline{G}_\text{T,NOCT}}\right)\left(\frac{\alpha_\text{P}\mu_\text{mp,STC}}{\tau\alpha}\right)} \\ \nonumber 
    +&\frac{\left(T_\text{c,NOCT}-T_\text{a,NOCT}\right)\left(\frac{\overline{G}_\text{T}\left(t\right)}{\overline{G}_\text{T,NOCT}}\right)\left[1-\frac{\mu_\text{mp,STC}\left(1-\alpha_\text{P}T_\text{c,STC}\right)}{\tau\alpha}\right]}{1+\left(T_\text{c,NOCT}-T_\text{a,NOCT}\right)\left(\frac{\overline{G}_\text{T}\left(t\right)}{\overline{G}_\text{T,NOCT}}\right)\left(\frac{\alpha_\text{P}\mu_\text{mp,STC}}{\tau\alpha}\right)},
\end{align}
where $T_\text{c,NOCT}$, $T_\text{a,NOCT}$, and $\overline{G}_\text{T,NOCT}$ are the nominal operating cell temperature (NOCT) of the PV panel, and the ambient temperature and solar radiation at which the NOCT is defined, respectively. Next, $\tau$, $\alpha$, and $\mu_\text{mp,STC}$ are the solar transmittance of any cover over the PV array and its solar absorptance, and the maximum power point efficiency of the PV panel under STC. This efficiency is equal to $\mu_\text{mp,STC}=\frac{P_\text{R,PV}}{a_\text{PV}\cdot b_\text{PV}\cdot\overline{G}_\text{T,STC}}$, where $a_\text{PV}$ and $b_\text{PV}$ are the dimensions of a single PV module. In turn, the parameter of $h_\text{PV}$ is the ground-relative altitude of the PV panels powering a specific cell.
\subsection{Wind Turbine}
\label{secWindTurbine}
Based on the contribution from \cite{Carrillo}, the output power in the time step $t$ of a single wind turbine under standard test conditions $\left(P_\text{WT,STC}\right)$ can be formulated as below: 
\begin{equation}
    \label{eqPowerWtSTC}
    P_\text{WT,STC}\left(t\right)=
    \begin{cases}
        F\big(v_\text{w}\left(t, h_\text{WT}\right)\big), & \text{if } v_\text{in} \leqslant v_\text{w}\left(t, h_\text{WT}\right) \leqslant v_\text{out} \\
        0, & \text{otherwise}
    \end{cases},
\end{equation}
where $h_\text{WT}$, $v_\text{w}\left(t, h_\text{WT}\right)$, and $F\left(v_\text{w}\left(t, h_\text{WT}\right)\right)$ are the ground-relative altitude of the WT, wind speed at the altitude equal to $h_\text{WT}$ at the current moment $t$, and the value of the power curve function (see \cite{WindTurbine}) of the WT for this wind speed, respectively. Conducting appropriate correction of $P_\text{WT,STC}$ parameter, it is possible to define the total power currently produced by the whole wind generators segment $\left(P_\text{WT}\right)$ of a certain cell \cite{HomerCalc}:
\begin{align}
    \label{eqPowerWT}
    P_\text{WT}\left(t\right)=N_\text{WT}\cdot P_\text{WT,STC}\left(t\right)\cdot\frac{\rho\left(t, h_\text{WT}\right)}{\rho_\text{STC}},
\end{align}
where $\rho_\text{STC}$ and $\rho\left(t, h_\text{WT}\right)$ are the air density at STC and the actual one at the level of $h_\text{WT}$ in the current time step $t$. Finally, $N_\text{WT}=N_\text{WT,s}\cdot N_\text{WT,p}$ is the total number of used wind turbines, where $N_\text{WT,s}$ and $N_\text{WT,p}$ are the numbers of WTs connected to each other in series and parallel, respectively.
\subsection{Battery System}
\label{secBatterySystem}
Each cell has its own battery system to receive the energy from the RES generators. This system is also the first power source to satisfy a cell's electrical demand. Only when the accumulators cannot provide any energy, it is delivered from the power lines grid. 
Inspired by \cite{Voltacon}, we propose a new model of energy management within the battery system $\left(E_\text{BATT}\right)$, which has been specified below:
\begin{equation}
    \label{eqBatteryEnergy}
    E_\text{BATT}\left(t\right)=
    \begin{cases}
        E_\text{BATT}\left(t'\right)+\Delta E_{\text{BATT},1}\left(t\right), & \text{if } \Delta E\left(t\right) > 0\\
        E_\text{BATT}\left(t'\right)+\Delta E_{\text{BATT},2}\left(t\right), & \text{otherwise}
    \end{cases},
\end{equation}
where $E_\text{BATT}\left(t'\right)$ is the energy handled by the battery system in the previous time step $t'$. Next, $\Delta E_{\text{BATT},1}$ and $\Delta E_{\text{BATT},2}$ are the energy amounts that have to be transferred from/to the battery system in the current time step $t$. Finally, $\Delta E$ is the energy balance, i.e., the difference between required energy and harvested one at the same moment. The parameters of $\Delta E_{\text{BATT},1}$ and $\Delta E_{\text{BATT},2}$ can be expressed by the formulas:
\begin{align}
    \label{eqBatteryEnergyDelta1}
    &\Delta E_{\text{BATT},1}\left(t\right)\\ \nonumber&=\text{max}\Big(\Delta E\left(t\right)\cdot\mu_\text{BATT},E_\text{BATT,max}-E_\text{BATT}\left(t'\right)\Big), 
\end{align}

\begin{align}
    \label{eqBatteryEnergyDelta2}
    \Delta E_{\text{BATT},2}\left(t\right)=\text{max}\left(\frac{\Delta E\left(t\right)}{\mu_\text{BATT}},-E_\text{BATT}\left(t'\right)\right),
\end{align}
where $\mu_\text{BATT}$ and $E_\text{BATT,max}$ are the efficiency of the used battery type and maximum energy the battery system is able to collect, respectively. The latter is equal to $E_\text{BATT,max}=N_\text{BATT}E'_\text{BATT,max}$, where $E'_\text{BATT,max}$ is the maximum energy of a single battery, and $N_\text{BATT}=N_\text{BATT,s}N_\text{BATT,p}$ is the total number of accumulator units in a battery system. The parameters of $N_\text{BATT,s}$ and $N_\text{BATT,p}$ are the numbers of batteries linked to each other in serial and parallel order, respectively. To evaluate the current energy balance $\left(\Delta E\right)$, the formula below was engaged:
\begin{align}
    \label{eqEnergyDelta}
    \Delta E\left(t\right)=\Bigg(P_\text{PV}\left(t\right)+P_\text{WT}\left(t\right)-\frac{P_\text{MIMO}\left(t\right)}{1-\sigma_\text{DC}}\Bigg)\Bigg(t-t'\Bigg), 
\end{align}
where $\sigma_\text{DC}$ is the loss factor related to DC supplying the hardware parts of the network cell.

\subsection{Atmospheric Parameters}
\label{secAtmoParams}
Finally, let us collect all auxiliary formulas used to calculate necessary atmospheric parameters. To approximate the actual speed of wind $\left(v_\text{w}\right)$ at the specific altitude $h$ in the time step $t$, the following mathematical equation can be used \cite{HomerCalc}:
\begin{align}
    \label{eqVw}
    v_\text{w}\left(t, h\right)=v_\text{w}\left(t, h_\text{WS}\right)\cdot\frac{\ln{\big(\left(h+h_\text{T}\right)/z_{0}\big)}}{\ln{\big(h_\text{WS}/z_{0}\big)}},
\end{align}
where $h_\text{WS}$, $h_\text{T}$, and $z_{0}$ are the absolute altitude, at which the measurement has been done (the altitude of the weather station -- WS), terrain altitude, and surface roughness length, respectively. The air density $\left(\rho\right)$ at the altitude $h$ and in the current time step $t$ can be calculated as follows \cite{Omni}:
\begin{align}
    \label{eqAirDensity}
    &\rho\left(t, h\right)=\\ \nonumber &\frac{p_\text{d}\left(t, h\right)}{R_\text{d}\cdot \big(T_\text{a}\left(t, h\right)+273.15\big)}+\frac{p_\text{v}\left(t, h\right)}{R_\text{v}\cdot \big(T_\text{a}\left(t, h\right)+273.15\big)},
\end{align}
where $R_\text{d}$ and $R_\text{v}$ are the specific gas constants for dry air and water vapor, respectively. Next, $p_\text{d}$ and $p_\text{v}$ are the pressures of dry air and water vapor. The latter at the altitude $h$ and in the time step $t$ can be expressed by the formula \cite{Omni}:
\begin{align}
    \label{eqVaporPressure}
    p_\text{v}\left(t, h\right)=6.1078\cdot 10^{\frac{7.5\cdot T_\text{a}\left(t, h\right)}{T_\text{a}\left(t, h\right)+237.3}},
\end{align}
The pressure of dry air at the same altitude and moment has been described by $p_\text{d}\left(t, h\right)=p\left(t, h\right)-p_\text{v}\left(t, h\right)$, 
where $p$ is the air pressure evaluated as \cite{Omni}: 
\begin{align}
    \label{eqPressureAtAltitude}
    p\left(t, h\right)=p_{0}\left(t\right)\cdot e^{\frac{-g\left(h\right)\cdot M\cdot \left(h+h_\text{T}-h_{0}\right)}{R\cdot T_\text{a}\left(t, h\right)}},
\end{align}
where $p_{0}$ is the air pressure at the reference level $h_{0}$. It was assumed that the reference level is the sea level altitude $\left(h_{0}=0\right.)$. Next referring to \cite{vCalc}, the gravitational acceleration is described by $g\left(h\right)=g_0\frac{r_\text{e}^{2}}{\left(r_\text{e}+h\right)^{2}}$, where $g_0$ and $r_\text{e}$ are the sea level acceleration and mean radius of the Earth, respectively. Finally, the formula to calculate the ambient temperature $\left(T_\text{a}\right)$ at the altitude $h$ and moment $t$ is shown below \cite{Omni}:
\begin{align}
    \label{eqTemperatureAtAltitude}
    T_\text{a}\left(t, h\right)=T_\text{a}(t, h_\text{WS})-0.0065\left(h+h_\text{T}-h_\text{WS}\right).
\end{align}

\section{Simulation Setup}
\label{secSimSetup}
The source code of the developed software was prepared in Java language. The examination of the system scenario described in Section \ref{secScenario} has been performed in the form of $10$ independent simulation runs each considering $4$ days of the previous year starting different seasons -- vernal equinox $\left(20^\text{th} \text{ March } 2022\right)$, summer solstice $\left(21^\text{st} \text{ June } 2022\right)$, autumn equinox $\left(23^\text{rd} \text{ September } 2022\right)$, and winter solstice $\left(21^\text{st} \text{ December } 2022\right)$. The parameters of users (location coordinates and traffic demand) have always been defined at the beginning of each simulation run. The assumed time step was equal to $1$ hour $\left(4\cdot24=96 \text{ steps per simulation run}\right)$, with which the weather data was updated, and then the calculations for energy prosumption were carried out. The simulation setup for network and energy designs is highlighted in Tab.~\ref{tabNetworkConfig} and \ref{tabProsumptionConfig}. 

\begin{table}[!h]
\centering
\caption{Network Design Configuration \cite{Castellanos, Björnson, AreaData, NetworkData}}
\label{tabNetworkConfig}
\resizebox{0.48\textwidth}{!}{
\begin{tabular}{|l|c|c|c|cccc|}
\hline
\multirow{3}{*}{}   & \multirow{3}{*}{Parameter} & \multirow{3}{*}{Sign} & \multirow{3}{*}{Unit} & \multicolumn{4}{c|}{Value}                                                                                                                                        \\ \cline{5-8} 
                    &                            &                       &                       & \multicolumn{3}{c|}{Cell of Base Station}                                                 & \multirow{2}{*}{\begin{tabular}[c]{@{}c@{}}Mobile \\ Station\end{tabular}} \\ \cline{5-7}
                    &                            &                       &                       & \multicolumn{1}{c|}{$1^\text{st}$}    & \multicolumn{1}{c|}{$2^\text{nd}$}   & \multicolumn{1}{c|}{$3^\text{rd}$}  &                                                                               \\ \hline
\multirow{4}{*}{\rotatebox[origin=c]{90}{Overall}} 
                    & Quantity                   & $K$            & --                     &                                    \multicolumn{3}{c|}{$8$}    & $300$                                                                      \\ \cline{2-8} 
                    & Movement Speed             & $v$            & {$\left[\text{m/s}\right]$}             & \multicolumn{3}{c|}{N/A}                                                          & $0$                                                                             \\ \cline{2-8} 
                    & Placement                  & --            & --                     & \multicolumn{3}{c|}{N/A}                                                          & outdoor                                               \\ \cline{2-8}
                    & Technology & -- & -- & \multicolumn{3}{c|}{$5\text{G}$} & N/A\\ 
                    \hline
\multirow{11}{*}{\rotatebox[origin=c]{90}{Band}}
                    & Frequency                  & $f$            & {$\left[\text{MHz}\right]$}             & \multicolumn{1}{c|}{$800$}  & \multicolumn{1}{c|}{$2100$} & \multicolumn{1}{c|}{$3500$} & N/A                                                                           \\ \cline{2-8} 
                    & Channel Bandwidth          & $B_\text{w}$            & {$\left[\text{MHz}\right]$}      & \multicolumn{1}{c|}{$80$}   & \multicolumn{1}{c|}{$120$}  & \multicolumn{1}{c|}{$120$}  & N/A                                                                           \\ \cline{2-8} 
                    & Used Subcarriers           & $N_\text{SC,u}$            & --                     & \multicolumn{3}{c|}{$320$}                                                          & N/A                                                                           \\ \cline{2-8} 
                    & Total Subcarriers          & $N_\text{SC,t}$            & --                     & \multicolumn{3}{c|}{$512$}                                                          & N/A                                                                           \\ \cline{2-8} 
                    & Sampling Factor            & SF                    & --                     & \multicolumn{3}{c|}{$1.536$}                                                        & N/A                                                                           \\ \cline{2-8} 
                    & Pilot Reuse Factor         & RF                    & --                     & \multicolumn{3}{c|}{$1$}                                                            & N/A                                                                           \\ \cline{2-8} 
                    & Coherence Time             & $t_\text{c}$            & {$\left[\text{ms}\right]$}              & \multicolumn{3}{c|}{$50$}                                                           & N/A                                                                           \\ \cline{2-8} 
                    & Coherence Bandwidth        & $B_\text{c}$            & {$\left[\text{MHz}\right]$}             & \multicolumn{3}{c|}{$1$}                                                            & N/A                                                                           \\ \cline{2-8} 
                    & TDD Duty Cycle DL          & $D_\text{DL}$            & {$\left[\text{\%}\right]$}              & \multicolumn{3}{c|}{$75$}                                                           & N/A                                                                           \\ \cline{2-8} 
                    & TDD Duty Cycle UL          & $D_\text{UL}$            & {$\left[\text{\%}\right]$}              & \multicolumn{3}{c|}{$25$}                                                           & N/A                                                                           \\ \cline{2-8} 
                    & Spatial Duty Cycle         & $S$            & {$\left[\text{\%}\right]$}              & \multicolumn{1}{c|}{$0$}    & \multicolumn{1}{c|}{$0$}    & \multicolumn{1}{c|}{$25$}   & N/A                                                                           \\ \hline
\multirow{6}{*}{\rotatebox[origin=c]{90}{Transceivers}}  & Antenna Height             & $h$            & {$\left[\text{m}\right]$}               & \multicolumn{3}{c|}{$\left(32, 46\right)$}                                                     & $1.5$                                                                           \\ \cline{2-8} 
                    & Antenna Elements           & $M$            & --                     & \multicolumn{1}{c|}{$1$}    & \multicolumn{1}{c|}{$1$}    & \multicolumn{1}{c|}{$64$}   & $1$                                                                             \\ \cline{2-8} 
                    & Antenna Gain               & $G_\text{a}$            & {$\left[\text{dBi}\right]$}             & \multicolumn{1}{c|}{$16$}   & \multicolumn{1}{c|}{$18$}   & \multicolumn{1}{c|}{$24$}   & $0$                                                                             \\ \cline{2-8} 
                    & Antenna Feeder Loss        & $L_\text{f}$            & {$\left[\text{dBi}\right]$}             & \multicolumn{1}{c|}{$2$}    & \multicolumn{1}{c|}{$2$}    & \multicolumn{1}{c|}{$3$}    & $0$                                                                             \\ \cline{2-8} 
                    & Max. Transmit Power        & $P_\text{TX, max}$            & {$\left[\text{dBm}\right]$}             & \multicolumn{1}{c|}{$46$}   & \multicolumn{1}{c|}{$49$}   & \multicolumn{1}{c|}{$53$}   & $23$                                                                            \\ \cline{2-8} 
                    & Noise Figure               & NF                    & {$\left[\text{dB}\right]$}              & \multicolumn{1}{c|}{$8$}    & \multicolumn{1}{c|}{$8$}    & \multicolumn{1}{c|}{$7$}    & N/A                                                                           \\ \hline
\multirow{7}{*}{\rotatebox[origin=c]{90}{Propagation}}  & Path Loss Model            & --                     & --                     & \multicolumn{3}{c|}{TR $38.901$}                                                    & N/A                                                                           \\ \cline{2-8} 
                    & Interference Margin        & IM                    & {$\left[\text{dB}\right]$}              & \multicolumn{3}{c|}{$2$}                                                            & $0$                                                                             \\ \cline{2-8} 
                    & Doppler Margin             & DM                    & {$\left[\text{dB}\right]$}              & \multicolumn{3}{c|}{$3$}                                                            & N/A                                                                           \\ \cline{2-8} 
                    & Fade Margin                & FM                    & {$\left[\text{dB}\right]$}              & \multicolumn{3}{c|}{$10$}                                                           & N/A                                                                           \\ \cline{2-8} 
                    & Shadow Margin              & SM                    & {$\left[\text{dB}\right]$}              & \multicolumn{1}{c|}{$12.8$} & \multicolumn{1}{c|}{$15.2$} & \multicolumn{1}{c|}{$10$}   & N/A                                                                           \\ \cline{2-8} 
                    & Implementation Loss        & IL                    & {$\left[\text{dB}\right]$}              & \multicolumn{1}{c|}{$0$}    & \multicolumn{1}{c|}{0}    & \multicolumn{1}{c|}{$3$}    & N/A                                                                           \\ \cline{2-8} 
                    & Soft Handover Gain         & $G_\text{SHO}$            & {$\left[\text{dB}\right]$}              & \multicolumn{3}{c|}{N/A}                                                          & $0$                                                                             \\ \hline
\end{tabular}}
\end{table}

\begin{table}[!t]
\centering
\caption{Energy Prosumption Configuration \cite{Björnson, Peng, Voltacon, Omni, vCalc, PvPanel, WindTurbine, Battery, AreaData, Arnold, Franklin, HomerGloss}}
\label{tabProsumptionConfig}
\resizebox{0.48\textwidth}{!}{
\begin{tabular}{|c|c|c|c|c|}
\hline
                    & Parameter                                                    & Sign                      & Unit       & Value                           \\ \hline
\multirow{15}{*}{\rotatebox[origin=c]{90}{Network Cell}}   & Fixed Power Component                                        & $P_\text{FIX}$                    & {$\left[\text{W}\right]$}    & $10$                              \\ \cline{2-5} 
                                 & Local Oscillator Power                                       & $P_\text{LO}$                     & {$\left[\text{W}\right]$}    & $0.2$                             \\ \cline{2-5} 
                                 & Circuit Components Power                                     & $P_\text{CC}$                     & {$\left[\text{W}\right]$}    & $0.4$                             \\ \cline{2-5} 
                                 & Encoding Power                                               & $P_\text{COD}$                    & {$\left[\text{W}\right]$}    & $0.1$                             \\ \cline{2-5} 
                                 & Decoding Power                                               & $P_\text{DEC}$                    & {$\left[\text{W}\right]$}    & $0.8$                             \\ \cline{2-5} 
                                 & Backhaul Traffic Power                                       & $P_\text{BT}$                     & {$\left[\text{W}\right]$}    & $0.25$                            \\ \cline{2-5} 
                                 & Auxiliary Power                                              & $P_\text{AUX}$                    & {$\left[\text{W}\right]$}    & $0$                               \\ \cline{2-5} 
                                 & Computational Efficiency                                     & $\eta_\text{BS}$                     & {$\left[\text{Gflops}/\text{W}\right]$}          & $75$                              \\ \cline{2-5}
                                 & Server Room (SR) Dimensions & $a_\text{s}$ x $b_\text{s}$ x $h_\text{s}$ & {$\left[\text{m}\right]$}    & $7$ x $5$ x $3.5$              \\ \cline{2-5} 
                                 & SR Target Temperature                               & $T_\text{s}$                      & {$\left[^\circ\text{C}\right]$}    & $18$                              \\ \cline{2-5} 
                                 & SR Heat Transfer Coeffi.                        & $\sigma_\text{s}$                  & --          & $2.037$                           \\ \cline{2-5} 
                                 & Circuit Heat Transfer Coeffi.                            & $\sigma_\text{CP}$                  & --          & $0.9$              \\ \cline{2-5}   
                                 & Cooling Loss Factor                                          & $\sigma_\text{cool}$               & --          & $0.1$           \\ \cline{2-5}
                                 & DC Loss Factor                                               & $\sigma_\text{DC}$                 & --          & $0.075$                           \\ \cline{2-5} 
                                 & Amplifier Efficiency                                         & $\mu_\text{PA}$                     & --          & $0.35$                                   \\ \hline
\multirow{18}{*}{\rotatebox[origin=c]{90}{PV Panels}}      & Model                                              & \multicolumn{3}{c|} {Solarland SLP$020$-$12$U}            \\ \cline{2-5}  
                                 & Nominal Voltage                           & $V_\text{n,PV}$                  & {$\left[\text{V}\right]$}    & $12$                              \\ \cline{2-5} 
                                 & Voltage at Max. Power                       & $V_\text{max,PV}$                & {$\left[\text{V}\right]$}    & $17.2$                            \\ \cline{2-5} 
                                 & Current at Max. Power                       & $I_\text{max, PV}$                & {$\left[\text{A}\right]$}    & $1.16$                            \\ \cline{2-5} 
                                 & Rated Power                               & $P_\text{R,PV}$                  & {$\left[\text{W}\right]$}    & $20$                              \\ \cline{2-5} 
                                 & Ground-relative Altitude                                      & $h_\text{PV}$                     & {$\left[\text{m}\right]$}    & $\left(27, 41\right)$                               \\ \cline{2-5}
                                 & Module Dimensions                 & $a_\text{PV}$ x $b_\text{PV}$             & {$\left[\text{m}\right]$}    & $0.576$ x $0.357$                   \\ \cline{2-5} 
                                 & Solar Radiation at STC                                       & $\overline{G}_\text{T,STC}$                    & {$\left[\text{W/}\text{m}^{2}\right]$}  & $1000$                            \\ \cline{2-5} 
                                 & Solar Radiation for NOCT                                      & $\overline{G}_\text{T,NOCT}$                   & {$\left[\text{W/}\text{m}^{2}\right]$}  & $800$                             \\ \cline{2-5}
                                 & Temperature under STC                                      & $T_\text{c,STC}$                    & {$\left[^\circ\text{C}\right]$}    & $25$                              \\ \cline{2-5} 
                                 & Temperature NOCT                                     & $T_\text{c,NOCT}$                   & {$\left[^\circ\text{C}\right]$}    & $47$                              \\ \cline{2-5} 
                                 & Temperature Coeffi. of Power                             & $\alpha_\text{PV}$                 & {$\left[\text{\%/}^\circ\text{C}\right]$} & $-0.5$                            \\ \cline{2-5} 
                                 & Solar Absorptance                                        & $\alpha$                    & --          & $0.3\sqrt{10}$                            \\ \cline{2-5}
                                 & Solar Transmittance                                          & $\tau$                    & --          & $0.3\sqrt{10}$                            \\ \cline{2-5}
                                 & Derating Factor                                          & $f_\text{PV}$                    & --          & $0.723$                            \\ \cline{2-5} 
                                 & Number in Serial Order                               & $N_\text{PV,s}$                  & --          & $4$                               \\ \cline{2-5} 
                                 & Number in Parallel Order                             & $N_\text{PV,p}$                  & --          & $16$                              \\ \cline{2-5} 
                                 & Total Number per Net. Cell                          & $N_\text{PV}$                     & --          & $64$                              \\ \hline
\multirow{14}{*}{\rotatebox[origin=c]{90}{Wind Turbine}}   & Model                              & \multicolumn{3}{c|} {Greatwatt S$1000$ $1200$W / $48$V}    \\ \cline{2-5}
                                 & Nominal Voltage                          & $V_\text{n,WT}$                   & {$\left[\text{V}\right]$}    & $48$                              \\ \cline{2-5} 
                                 & Rated Power                              & $P_\text{R,WT}$                   & {$\left[\text{W}\right]$}    & $1000$                            \\ \cline{2-5}  
                                 & Rated Wind Speed                                             & $v_\text{r}$                      & {$\left[\text{m/s}\right]$}  & $10$                              \\ \cline{2-5} 
                                 & Cut-In Wind Speed                                            & $v_\text{in}$                     & {$\left[\text{m/s}\right]$}  & $3$                               \\ \cline{2-5} 
                                 & Cut-Out Wind Speed                                           & $v_\text{out}$                    & {$\left[\text{m/s}\right]$}  & $16.2$                            \\ \cline{2-5}
                                 & Rotor Radius                              & $r_\text{WT}$                     & {$\left[\text{m}\right]$}    & $1.09$                            \\ \cline{2-5} 
                                 & Ground-relative Altitude                                      & $h_\text{WT}$                     & {$\left[\text{m}\right]$}    & $\left(35, 49\right)$                               \\ \cline{2-5}  
                                 & Surface Roughness Length                                     & $z_\text{0}$                      & {$\left[\text{m}\right]$}          & $3.0$               \\ \cline{2-5}
                                 & Air Density at STC                                           & $\rho_\text{STC}$                  & {$\left[\text{kg/}\text{m}^{3}\right]$} & $1.225$                           \\ \cline{2-5}
                                 & Number of Rotor Blades                                   & $i_\text{WT}$                     & --          & $3$                               \\ \cline{2-5} 
                                 & Number in Serial Order                            & $N_\text{WT,s}$                   & --          & $1$                               \\ \cline{2-5} 
                                 & Number in Parallel Order                          & $N_\text{WT,p}$                   & --          & $1$                               \\ \cline{2-5} 
                                 & Total Number per Net. Cell                       & $N_\text{WT}$                     & --          & $1$                                
                                  \\ \hline
\multirow{16}{*}{\rotatebox[origin=c]{90}{Battery System}} & Model                                                & \multicolumn{3}{c|} {Solise Battery $48$V $60$Ah $\text{LiFePO}_4$} \\ \cline{2-5} 
                                 & Nominal Voltage                          & $V_\text{n,BATT}$                & {$\left[\text{V}\right]$}    & $51.2$                            \\ \cline{2-5} 
                                 & Charging Voltage                         & $V_\text{c,BATT}$                & {$\left[\text{V}\right]$}    & $57.6$                            \\ \cline{2-5} 
                                 & Discharging Voltage                      & $V_\text{d,BATT}$                & {$\left[\text{V}\right]$}    & $51.2$                            \\ \cline{2-5} 
                                 & Charging Current                         & $I_\text{c,BATT}$                & {$\left[\text{A}\right]$}    & $30$                              \\ \cline{2-5} 
                                 & Rapid Charging Current                   & $I_\text{r,BATT}$                & {$\left[\text{A}\right]$}    & $50$                              \\ \cline{2-5} 
                                 & Discharging Current                      & $I_\text{d,BATT}$                & {$\left[\text{A}\right]$}    & $50$                              \\ \cline{2-5} 
                                 & Capacity                                 & $C_\text{BATT}$                   & {$\left[\text{Ah}\right]$}   & $60.78$                           \\ \cline{2-5} 
                                 & Provided Energy                          & $E'_\text{BATT,max}$                   & {$\left[\text{Wh}\right]$}   & $3112$                            \\ \cline{2-5} 
                                 & Max. Depth of Discharge                  & $\text{DoD}_\text{max}$                  & {$\left[\text{\%}\right]$}   & $100$                             \\ \cline{2-5} 
                                 & Primary State of Charge                  & $\text{SoC}_\text{p}$                    & {$\left[\text{\%}\right]$}   & $100$                             \\ \cline{2-5} 
                                 & Battery's Efficiency                                 & $\mu_\text{BATT}$                   & --          & $0.95$                        \\ \cline{2-5}    
                                 & Number of Cycles                         & $N_\text{BC}$                     & --          & $2000$                            \\ \cline{2-5} 
                                 & Number in Serial Order                                & $N_\text{BATT,s}$                 & --          & $1$                               \\ \cline{2-5} 
                                 & Number in Parallel Order                             & $N_\text{BATT,p}$                 & --          & $6$                               \\ \cline{2-5}
                                 & Total Number per Net. Cell                          & $N_\text{BATT}$                   & --          & $6$                               \\ \hline 
\multirow{10}{*}{\rotatebox[origin=c]{90}{Other}} & Reference Altitude                                        & $h_{0}$                    & {$\left[\text{m}\right]$}    & $0$                              \\ \cline{2-5} 
                                 & Terrain Absolute Altitude                                       & $h_\text{T}$                     & {$\left[\text{m}\right]$}    & $54.44$                             \\ \cline{2-5} 
                                 & Weather Station Absolute Alti.                                     & $h_\text{WS}$                     & {$\left[\text{m}\right]$}    & $90$                             \\ \cline{2-5}
                                 & Buildings Relative Altitude                                      & $h_\text{BU}$                     & {$\left[\text{m}\right]$}    & $\left(27, 41\right)$                               \\ \cline{2-5}
                                 & Mean Radius of the Earth                                     & $r_\text{e}$                     & {$\left[\text{m}\right]$}    & $6371009$                             \\ \cline{2-5}
                                 & Sea Level Gravitational Accel.                                     & $g_{0}$                     & {$\left[\text{m}/\text{s}^{2}\right]$}    & $9.80665$                             \\ \cline{2-5}
                                 & Air Molar Mass                                               & $m_\text{air}$                    & {$\left[\text{kg}/\text{mol}\right]$}    & $0.0289644$                             \\ \cline{2-5}
                                 & Universal Gas Constant                                     & $R_\text{u}$                     & {$\left[\frac{\text{N}\cdot \text{m}}{\text{mol}\cdot \text{K}}\right]$}    & $8.31432$                             \\ \cline{2-5} 
                                 & Dry Air Gas Constant                                     & $R_\text{d}$                     & {$\left[\text{J}/\left(\text{kg}\cdot\text{K}\right)\right]$}    & $287.058$                             \\ \cline{2-5} 
                                 & Water Vapor Gas Constant                                     & $R_\text{v}$                     & {$\left[\text{J}/\left(\text{kg}\cdot\text{K}\right)\right]$}    & $461.495$                             \\ \hline
\end{tabular}}
\end{table}

\begin{figure}[!t]
\centering
\includegraphics[width=0.465\textwidth]{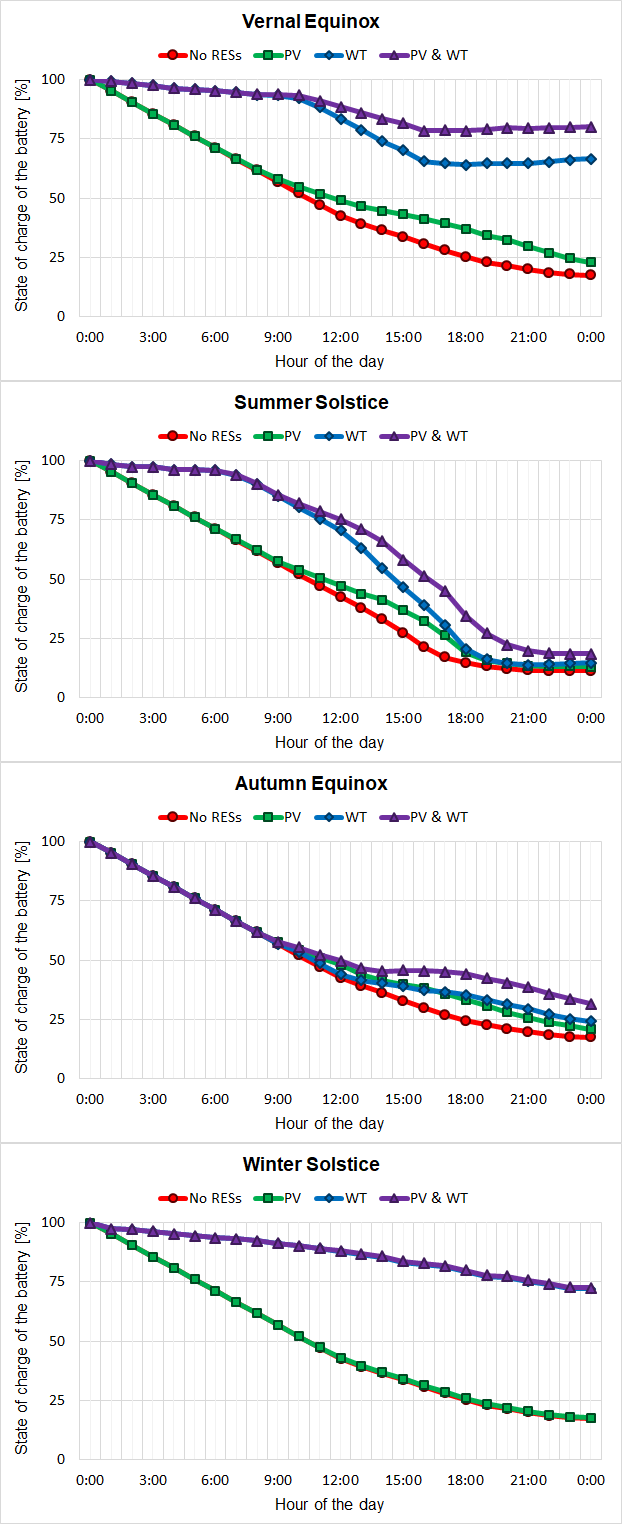}
\caption{Average state of charge of the battery in different seasons of the year.}
\label{figBatterySocs}
\end{figure}

\section{Results}
\label{secResults}
The results shown in Fig.~\ref{figBatterySocs} present achieved characteristics for battery system behavior according to the season of the year. The attached plots were prepared by averaging the state of charge of the battery systems of all cells over a specific time step from each simulation run. Based on the same data, the impact of using PV panels and wind turbines on extending the lifetime of a cell's battery system and reducing energy consumption from conventional sources was included in Tab.~\ref{tabEnergyRES}. By engaging RES generators within the wireless system, we were able to gain additional power support for the battery banks for each considered date. However, it can be noticed that much more effective type of RES generator (of the two presented) for the area specified by the weather conditions similar to the ones observed in Poland is the wind turbine. The most effective season from the perspective of using WTs is the winter solstice. On that day, the energy consumption from the electrical grid based on fossil fuels was reduced by about $74.24\%$. This in turn extended the lifetime of each battery on average by about $195.11\%$ that day. As the second best case was the vernal equinox reducing the use of conventional resources and extending the battery lifetime by about $64.74\%$ and $147.56\%$, respectively. Other dates were characterized by unfavorable wind conditions, i.e., too high (summer solstice) or too low (autumn equinox) speed of it for most of the day. Although the most solar energy resources were obtained during the summer solstice, the vernal equinox was much more efficient for equipping cells with PV panels $\left(\text{AEBL: }7.24\% \text{, AREC: } 15.65\%\right)$. The reason for that was the occurrence of additional power consumption caused by the need to cool the hardware in server rooms during high summer temperatures. For the rest of the dates, recorded solar radiation density values were at a very low level (especially for the winter solstice, where there are almost no energy savings). Furthermore, for the vernal equinox, the combination of both energy generator types was remarked to be the best solution in terms of the efficiency of supporting the existing grid of power lines $\left(\text{AEBL: }313.32\% \text{, AREC: } 76.63\%\right)$. Successively in the classification were the winter solstice $\left(\text{AEBL: }200.24\% \text{, AREC: } 74.66\%\right)$, autumn equinox (more efficient in AEBL than summer solstice), and summer solstice (more efficient in AREC than autumn equinox). From the perspective of the whole year, the most efficient solution was the deployment of both types of RES generators $\left(\text{AEBL: }70.24\%\text{, AREC: }50.69\%\right)$. 

\begin{table}[!htb]
\caption{Energy characteristics for various RES configurations}
\label{tabEnergyRES}
\resizebox{0.485\textwidth}{!}{
\begin{tabular}{|c|ccc|}
\hline
\rowcolor[HTML]{FFFFFF} 
\cellcolor[HTML]{FFFFFF}{\color[HTML]{000000} }                   & \multicolumn{3}{c|}{\cellcolor[HTML]{FFFFFF}{\color[HTML]{000000} Total (and peak) energy obtained per cell {$\left[\text{kWh}\right]$}}}                                                                                               \\ \cline{2-4} 
\rowcolor[HTML]{FFFFFF} 
\multirow{-2}{*}{\cellcolor[HTML]{FFFFFF}{\color[HTML]{000000} }} & \multicolumn{1}{c|}{\cellcolor[HTML]{FFFFFF}{\color[HTML]{000000} \textit{PV Panels}}}  & \multicolumn{1}{c|}{\cellcolor[HTML]{FFFFFF}{\color[HTML]{000000} \textit{Wind Turbine}}} & {\color[HTML]{000000} \textit{PV \& WT}}  \\ \hline
\rowcolor[HTML]{FFFFFF} 
{\color[HTML]{000000} Vernal Equinox}                             & \multicolumn{1}{c|}{\cellcolor[HTML]{FFFFFF}{\color[HTML]{000000} $91.2 \left(0.48\right)$}}         & \multicolumn{1}{c|}{\cellcolor[HTML]{FFFFFF}{\color[HTML]{000000} $446.5 \left(1.27\right)$}}          & {\color[HTML]{000000} {\ul $537.7 \left(1.56\right)$}} \\ \hline
\rowcolor[HTML]{FFFFFF} 
{\color[HTML]{000000} Summer Solstice}                            & \multicolumn{1}{c|}{\cellcolor[HTML]{FFFFFF}{\color[HTML]{000000} {\ul $109.91 \left(0.73\right)$}}} & \multicolumn{1}{c|}{\cellcolor[HTML]{FFFFFF}{\color[HTML]{000000} $237.92 \left(1.21\right)$}}         & {\color[HTML]{000000} $347.83 \left(1.39\right)$}      \\ \hline
\rowcolor[HTML]{FFFFFF} 
{\color[HTML]{000000} Autumn Equinox}                             & \multicolumn{1}{c|}{\cellcolor[HTML]{FFFFFF}{\color[HTML]{000000} $67.09 \left(0.52\right)$}}        & \multicolumn{1}{c|}{\cellcolor[HTML]{FFFFFF}{\color[HTML]{000000} $107.54 \left(0.87\right)$}}         & {\color[HTML]{000000} $174.63 \left(1.29\right)$}      \\ \hline
\rowcolor[HTML]{FFFFFF} 
{\color[HTML]{000000} Winter Solstice}                            & \multicolumn{1}{c|}{\cellcolor[HTML]{FFFFFF}{\color[HTML]{000000} $3.39 \left(0.03\right)$}}         & \multicolumn{1}{c|}{\cellcolor[HTML]{FFFFFF}{\color[HTML]{000000} {\ul $460.91 \left(1.11\right)$}}}   & {\color[HTML]{000000} $464.31 \left(1.11\right)$}      \\ \hline
\rowcolor[HTML]{6665CD} 
{\color[HTML]{FFFFFF} Annual average}                             & \multicolumn{1}{c|}{\cellcolor[HTML]{6665CD}{\color[HTML]{FFFFFF} $24,783.63$}}           & \multicolumn{1}{c|}{\cellcolor[HTML]{6665CD}{\color[HTML]{FFFFFF} $114,324$}}               & {\color[HTML]{FFFFFF} $139,107.6$}          \\ \hline
\end{tabular}
}
\\ \\ \\
\resizebox{0.48\textwidth}{!}{
\begin{tabular}{|
>{\columncolor[HTML]{FFFFFF}}c |
>{\columncolor[HTML]{FFFFFF}}c 
>{\columncolor[HTML]{FFFFFF}}c 
>{\columncolor[HTML]{FFFFFF}}c 
>{\columncolor[HTML]{FFFFFF}}c |}
\hline
\cellcolor[HTML]{FFFFFF}{\color[HTML]{000000} }                   & \multicolumn{4}{c|}{\cellcolor[HTML]{FFFFFF}{\color[HTML]{000000} Average extension of battery lifetime (AEBL) {$\left[\text{\%}\right]$}}}                                                                                                                                                                                                   \\ \cline{2-5} 
\multirow{-2}{*}{\cellcolor[HTML]{FFFFFF}{\color[HTML]{000000} }} & \multicolumn{1}{c|}{\cellcolor[HTML]{FFFFFF}{\color[HTML]{000000} \textit{No RESs}}} & \multicolumn{1}{c|}{\cellcolor[HTML]{FFFFFF}{\color[HTML]{000000} \textit{PV Panels}}} & \multicolumn{1}{c|}{\cellcolor[HTML]{FFFFFF}{\color[HTML]{000000} \textit{Wind Turbine}}} & {\color[HTML]{000000} \textit{PV \& WT}} \\ \hline
{\color[HTML]{000000} Vernal Equinox}                             & \multicolumn{1}{c|}{\cellcolor[HTML]{FFFFFF}{\color[HTML]{000000} $0$}}                & \multicolumn{1}{c|}{\cellcolor[HTML]{FFFFFF}{\color[HTML]{000000} {\ul $7.24$}}}         & \multicolumn{1}{c|}{\cellcolor[HTML]{FFFFFF}{\color[HTML]{000000} $147.56$}}                & {\color[HTML]{000000} {\ul $313.32$}}      \\ \hline
{\color[HTML]{000000} Summer Solstice}                            & \multicolumn{1}{c|}{\cellcolor[HTML]{FFFFFF}{\color[HTML]{000000} $0$}}                & \multicolumn{1}{c|}{\cellcolor[HTML]{FFFFFF}{\color[HTML]{000000} $1.96$}}               & \multicolumn{1}{c|}{\cellcolor[HTML]{FFFFFF}{\color[HTML]{000000} $3.94$}}                  & {\color[HTML]{000000} $8.62$}              \\ \hline
{\color[HTML]{000000} Autumn Equinox}                             & \multicolumn{1}{c|}{\cellcolor[HTML]{FFFFFF}{\color[HTML]{000000} $0$}}                & \multicolumn{1}{c|}{\cellcolor[HTML]{FFFFFF}{\color[HTML]{000000} $4.28$}}               & \multicolumn{1}{c|}{\cellcolor[HTML]{FFFFFF}{\color[HTML]{000000} $8.76$}}                  & {\color[HTML]{000000} $20.54$}             \\ \hline
{\color[HTML]{000000} Winter Solstice}                            & \multicolumn{1}{c|}{\cellcolor[HTML]{FFFFFF}{\color[HTML]{000000} $0$}}                & \multicolumn{1}{c|}{\cellcolor[HTML]{FFFFFF}{\color[HTML]{000000} $0.22$}}               & \multicolumn{1}{c|}{\cellcolor[HTML]{FFFFFF}{\color[HTML]{000000} {\ul $195.11$}}}          & {\color[HTML]{000000} $200.24$}            \\ \hline
\cellcolor[HTML]{6665CD}{\color[HTML]{FFFFFF} Annual average}                    & \multicolumn{1}{c|}{\cellcolor[HTML]{6665CD}{\color[HTML]{FFFFFF} $0$}}       & \multicolumn{1}{c|}{\cellcolor[HTML]{6665CD}{\color[HTML]{FFFFFF} $3.34$}}      & \multicolumn{1}{c|}{\cellcolor[HTML]{6665CD}{\color[HTML]{FFFFFF} $51.15$}}        & \cellcolor[HTML]{6665CD}{\color[HTML]{FFFFFF} $70.24$}    \\ \hline
\end{tabular}}
\\ \\ \\
\resizebox{0.48\textwidth}{!}{
\begin{tabular}{|
>{\columncolor[HTML]{FFFFFF}}c |
>{\columncolor[HTML]{FFFFFF}}c 
>{\columncolor[HTML]{FFFFFF}}c 
>{\columncolor[HTML]{FFFFFF}}c 
>{\columncolor[HTML]{FFFFFF}}c |}
\hline
\cellcolor[HTML]{FFFFFF}{\color[HTML]{000000} }                   & \multicolumn{4}{c|}{\cellcolor[HTML]{FFFFFF}{\color[HTML]{000000} Average reduction in energy consumption (AREC) {$\left[\text{\%}\right]$}}}                                                                                                                                                                                                   \\ \cline{2-5} 
\multirow{-2}{*}{\cellcolor[HTML]{FFFFFF}{\color[HTML]{000000} }} & \multicolumn{1}{c|}{\cellcolor[HTML]{FFFFFF}{\color[HTML]{000000} \textit{No RESs}}} & \multicolumn{1}{c|}{\cellcolor[HTML]{FFFFFF}{\color[HTML]{000000} \textit{PV Panels}}} & \multicolumn{1}{c|}{\cellcolor[HTML]{FFFFFF}{\color[HTML]{000000} \textit{Wind Turbine}}} & {\color[HTML]{000000} \textit{PV \& WT}} \\ \hline
{\color[HTML]{000000} Vernal Equinox}                             & \multicolumn{1}{c|}{\cellcolor[HTML]{FFFFFF}{\color[HTML]{000000} $0$}}                & \multicolumn{1}{c|}{\cellcolor[HTML]{FFFFFF}{\color[HTML]{000000} {\ul $15.65$}}}         & \multicolumn{1}{c|}{\cellcolor[HTML]{FFFFFF}{\color[HTML]{000000} $64.74$}}                & {\color[HTML]{000000} {\ul $76.63$}}      \\ \hline
{\color[HTML]{000000} Summer Solstice}                            & \multicolumn{1}{c|}{\cellcolor[HTML]{FFFFFF}{\color[HTML]{000000} $0$}}                & \multicolumn{1}{c|}{\cellcolor[HTML]{FFFFFF}{\color[HTML]{000000} $12.33$}}               & \multicolumn{1}{c|}{\cellcolor[HTML]{FFFFFF}{\color[HTML]{000000} $22.97$}}                  & {\color[HTML]{000000} $35.09$}              \\ \hline
{\color[HTML]{000000} Autumn Equinox}                             & \multicolumn{1}{c|}{\cellcolor[HTML]{FFFFFF}{\color[HTML]{000000} $0$}}                & \multicolumn{1}{c|}{\cellcolor[HTML]{FFFFFF}{\color[HTML]{000000} $10.93$}}               & \multicolumn{1}{c|}{\cellcolor[HTML]{FFFFFF}{\color[HTML]{000000} $17.33$}}                  & {\color[HTML]{000000} $27.52$}             \\ \hline
{\color[HTML]{000000} Winter Solstice}                            & \multicolumn{1}{c|}{\cellcolor[HTML]{FFFFFF}{\color[HTML]{000000} $0$}}                & \multicolumn{1}{c|}{\cellcolor[HTML]{FFFFFF}{\color[HTML]{000000} $0.64$}}               & \multicolumn{1}{c|}{\cellcolor[HTML]{FFFFFF}{\color[HTML]{000000} {\ul $74.24$}}}          & {\color[HTML]{000000} $74.66$}            \\ \hline
\cellcolor[HTML]{6665CD}{\color[HTML]{FFFFFF} Annual average}                    & \multicolumn{1}{c|}{\cellcolor[HTML]{6665CD}{\color[HTML]{FFFFFF} $0$}}       & \multicolumn{1}{c|}{\cellcolor[HTML]{6665CD}{\color[HTML]{FFFFFF} $10.23$}}      & \multicolumn{1}{c|}{\cellcolor[HTML]{6665CD}{\color[HTML]{FFFFFF} $41.55$}}        & \cellcolor[HTML]{6665CD}{\color[HTML]{FFFFFF} $50.69$}    \\ \hline
\end{tabular}}
\end{table}

\section{Conclusion}
\label{secConclusion}
The contribution presented in this paper highlights the advantages related to the use of PV panels and wind turbines as power generators in $5$G cellular networks. For the considered scenario, a very significant accomplishment has been observed. The power savings (and the resulting financial ones) exceeded the level of $50\%$ per year in comparison to the case, in which the system is supplied only by the conventional energy grid. Although RESs are characterized by time-varying harvesting processes, by appropriate management of available resources (radio and energy) using optimizing algorithms (e.g., traffic steering, resource allocation, etc.) we are able to improve already achieved results or even ensure energy autonomy for cellular network without worsening the quality of mobile services delivered to users. However, the implementation of those algorithms will be taken into account in future work.


\section*{Acknowledgment}
The cooperation between the authors has been initiated within the COST CA$10210$ INTERACT. M.~Deruyck is a Post-Doctoral Fellow of the FWO-V (Research Foundation -- Flanders, ref: $12\text{Z}5621\text{N}$). The work by A. Samorzewski and A. Kliks was realized within project no. $2021/43/\text{B}/\text{ST}7/01365$ funded by National Science Center in Poland. 

\end{document}